# Synergistic Astrophysics in the Ultraviolet using Active Galactic Nuclei

A Response to NASA/SMD Request For Information (RFI) on Science Objectives and Requirements for the Next NASA UV/Visible Mission Concepts (NNH12ZDA008L)

Gerard Kriss (STScI) Lead Submitter. Contacts: gak@stsci.edu, 410-338-4353.
Nahum Arav (Virginia Tech; arav@vt.edu), Anton Koekemoer (STScI; koekemoer@stsci.edu),
Smita Mathur (Ohio State; smita@astronomy.ohio-state.edu), Bradley M. Peterson (Ohio State; peterson@astronomy.ohio-state.edu) Jennifer E. Scott (Towson University; jescott@towson.edu)

*Abstract*: Observing programs comprising multiple scientific objectives will enhance the productivity of NASA's next UV/Visible mission. Studying active galactic nuclei (AGN) is intrinsically important for understanding how black holes accrete matter, grow through cosmic time, and influence their host galaxies. At the same time, the bright UV continuum of AGN serves as an ideal background light source for studying foreground gas in the intergalactic medium (IGM), the circumgalactic medium (CGM) of individual galaxies, and the interstellar medium (ISM) and halo of the Milky Way. A well chosen sample of AGN can serve as the observational backbone for multiple spectroscopic investigations including quantitative measurements of outflows from AGN, the structure of their accretion disks, and the mass of the central black hole.

Understanding how black holes accrete matter, grow through cosmic time, and influence their host galaxies is crucial for our understanding of galaxy evolution. Outflows from AGN, visible as blue-shifted ultraviolet and X-ray absorption lines from highly ionized species (Crenshaw et al. 2003), may be at the heart of feedback processes that regulate the growth of the host galaxy and chemically enrich its surroundings. The energy and momentum of outflowing winds from AGN expel gas from the host galaxy and inhibit star formation. In terms of color-magnitude diagrams, this ultimately moves AGN from the "Blue Cloud" across the "Green Valley" and onto the "Red Sequence" (Baldry et al. 2004). Shutting down further star formation limits galaxy growth, which is necessary to produce the observed galaxy luminosity function (Cole et al. 2001, Huang et al. 2003). The end result of AGN feedback couples black hole growth to galaxy growth, leading to the observed correlation between the mass of the black hole and the velocity dispersion of the spheroid of the host galaxy ($M_{BH}$-$\sigma$) (Di Matteo et al. 2005).

The central power source for AGN is accretion onto the central black hole through a luminous accretion disk. Most AGN emit their energy in the far and extreme ultraviolet energy range with a peak at ~1200 Å, extending into the extreme ultraviolet (Telfer et al. 2002; Shang et al. 2011). Temperatures forming such a peak are too cool for thermal radiation from the accretion disk to continue to the soft X-ray band (e.g., Done et al. 2012), and a likely explanation for the extreme ultraviolet continuum is Comptonization of the disk spectrum by a warm, ionized coronal layer just above the disk or near its inner edge. Direct observation of this portion of the spectrum in intermediate redshift AGN (z~1) and correlation with the longer-wavelength thermal continuum to study time lags associated with the Comptonized reprocessing would enable us to assess the geometry of the accretion disk.

The inferences summarized above have been gleaned from UV and optical observations of a few dozen, mostly local (z < 0.15), AGN. While current observations have enabled us to produce a general picture of AGN structure and how feedback might influence galaxy formation, we still lack firm quantification of accretion disk structure, the mass and energy flux in outflows, and the abundances in the outflowing gas. Models of AGN feedback usually include these inputs as parametric entries with little microphysics motivating the choices. Measuring such quantities in large samples of AGN over a range of redshift, luminosity and environment are necessary both to test models of galaxy formation, and supply the necessary physics. An 8-m UV telescope and spectrograph sensitive from 900—3200 Å could give sensitivity of 100x that of the Cosmic Origins Spectrograph (COS) on the Hubble Space Telescope (HST), enabling observations of a sample an order of magnitude or more greater than the few AGN currently observed in detail.

In this paper we present some sample scientific programs that could be enabled by such a dramatic increase in sensitivity. Furthermore, each of these programs could be accomplished with well-chosen samples of AGN and observations that simultaneously satisfy the scientific objectives of other compelling scientific investigations.

## Quantifying Outflows in Nearby AGN

An AGN outflow with a kinetic luminosity of 0.5% (Hopkins & Elvis 2010) to 5% (Di Matteo et al. 2005) of the Eddington luminosity of the black hole provides sufficient feedback to couple black hole growth to the evolution of the host galaxy. Measuring the kinetic luminosity of an AGN wind, however, is difficult. Assuming the outflow is in the form of a partial thin spherical shell moving with velocity v, its mass flux, $\dot{M}$, and kinetic luminosity, $\dot{E}_k$, are given by:

$$\dot{M} = 4\pi \Delta\Omega R N_H \mu m_p v$$
$$\dot{E}_k = \tfrac{1}{2}\dot{M}v^2$$

where $\Delta\Omega$ is the fraction of the total solid angle occupied by the outflow, $R$ is the distance of the outflow from the central source, $N_H$ is the total hydrogen column density of the outflow, $m_p$ is the mass of the proton, and $\mu=1.4$ is the molecular weight of the plasma per proton. Observations of UV absorption lines are the key to measuring all these necessary quantities.

For nearby AGN, the 900—3200 Å band contains key spectral diagnostics that let us measure the kinematics and abundance of highly ionized gas in the vicinity of the black hole. In particular, this wavelength range includes the Lyman lines of neutral hydrogen, and the lithium-like doublets of O VI, N V, Si IV, and C IV. High signal-to-noise observations of the absorption troughs of the Lyman series and these doublets allow us to measure ionic column densities and covering fractions as a function of outflow velocity in a model-independent way (Hamann et al. 1991; Arav et al. 1999). Combining ionic column densities with photoionization models yields the total column density and the ionization parameter. (An important additional product of the photoionization models is the absolute abundances of the elements in the outflow [Arav et al. 2007].) However, to determine the kinetic luminosity, we also need to determine the distance, R, of the outflow, which is linked to the gas density via the ionization parameter, $\xi = L_{ion} / nR^2$.

Measuring the gas density is the most difficult part of the observational problem. Density-sensitive transitions are one approach, but at low redshift, only low-ionization transitions of C II, C III, and Fe II are available. Since higher-ionization gas dominates the mass flux in the AGN winds (Crenshaw & Kraemer 2012), these are of limited utility. Nevertheless, the C III* λ1176 transitions have been used to establish the distance of outflow components in a few AGN (NGC 4151, Kraemer et al. 2006; NGC 3783, Gabel et al. 2005). A more generally applicable approach has been to monitor changes in the absorption components in the outflows and measure the timescale of their response to changes in the ionizing continuum, as shown in Fig. 1. The ionization/recombination timescales then give the density of the absorbing gas. Since this requires repeated high S/N observations, the method has been successfully applied to only the nearest and brightest AGN. Crenshaw & Kraemer (2012) give a summary of the best results for both techniques that comprises a total of only 10 objects.

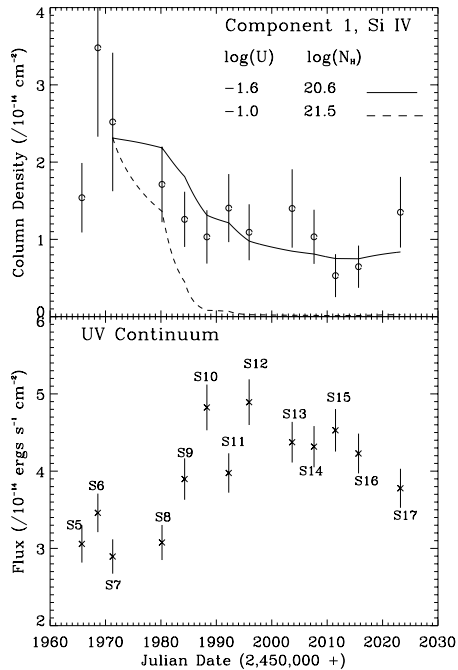

**Figure 1:** (Top) Variations in Si IV column density during the monitoring observations of NGC 3783 (Gabel et al. 2005). The solid line shows the best-fit solution based on simultaneously matching the low and high-state measurements. The dashed line shows a higher-ionization solution that fits only the low-state column densities. (Bottom) The UV continuum light curve is shown for comparison.

To make significant progress in quantifying AGN outflows in the local universe (z<0.15) requires the following:

1. High sensitivity (100x COS) covering 900—3200 Å to enable high S/N observations of outflow absorption signatures in the Lyman lines, O VI, N V, Si IV, and C IV. This would enable surveys of hundreds of AGN, which could be accomplished using the same sample of background sources used to probe the circum-galactic medium of intervening galaxies.
2. The same high sensitivity would enable repeated observations of a select subsample of AGN to measure the ionization response of the absorbers and thereby measure the density and distance of the absorbing gas. Such repeated observations could be part of a reverberation-mapping program that mapped the two-dimensional kinematics of the broad-line region in these same AGN. (See the white paper on reverberation mapping by Peterson et al. 2012.)

## Outflows in AGN at Intermediate Redshift

At intermediate redshifts, 0.2 < z < 2.0, extreme ultraviolet absorption lines of other highly ionized species become visible in the 900—3200 Å band. Ne VIII λλ770,780, Mg X λλ610,625 and Si XII λλ499,521 probe gas at ionization levels comparable to the O VII and O VIII features commonly seen in X-rays from local AGN. These ions have ionization potentials comparable to the X-ray absorbing gas detected in O VII and O VIII in bright, local AGN (which dominate the mass and kinetic energy flux, Crenshaw & Kraemer 2012). These UV ions have currently only been detected in the brightest intermediate-redshift AGN (Telfer et al. 1998; Muzahid et al. 2012). At intermediate redshifts, higher-ionization density-sensitive lines are redshifted into the

UV. Pairs of density-sensitive, ground+excited transitions of O III, O IV, O V, and Si IV become visible above redshifts of a few tenths, and they enable the direct measurement of the density in high-ionization gas in a single observation, as demonstrated by Arav et al. (2012) (see Fig. 2).

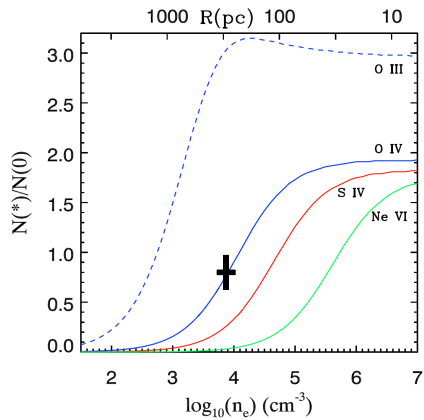

**Figure 2:** Outflow distance diagnostics for high ionization outflows. We first determine the electron number density ($n_e$) by measuring the column density ratio between the excited and ground state energy level of a given ion. The expected curves are shown here for 4 ionic species. For component A in HE0238–1904, the measured value of OIV*/OIV yields $n_e = 10^{3.9}$ cm$^{-3}$. Photoionization models yield the ionization parameter $\xi$, total column density $N_H$, and $n_e = 1.2 n_H$. Therefore, from the definition of $\xi$ we obtain R ∼ 700 pc. The top X-axis gives the distances for HE0238-1904, easily scalable for other objects.

At similar intermediate redshifts, X-ray diagnostic lines such as O VII and O VIII are absorbed by the local ISM, and X-ray fluxes are too low for spectroscopy. This makes studying the evolution of outflows difficult in the X-ray. In the UV at z<2, the integrated Lyα forest and continuum has an opacity of <10% (Zheng et al. 1997). An instrument with sensitivities of 100x COS would enable the detailed study of these UV species in hundreds of AGN. A UV spectrograph would be far more sensitive to outflows dominated by warm absorbers than any proposed future X-ray telescope, and with resolving power of 20,000, it would enable detailed kinematical studies as well. Again, the same sample of AGN used as background sources for probing the ISM, the IGM and the CGM would provide a sample enabling us to characterize the evolution of AGN outflows with redshift. This span of time covers the evolution of the cosmic star formation rate from its peak at z=2 until the present (e.g., Hopkins & Beacom 2006). Since outflows may be a key ingredient in regulating star formation and galaxy growth, understanding their co-evolution over the same time interval is crucial.

## The Physics of the Accretion Disk Spectrum in the Extreme UV

AGN spectral energy distributions peak in the ultraviolet (Elvis et al. 1994; Telfer et al. 2002; Shang et al. 2011). While the bulk of this emission is likely the thermal emission from an optically thick accretion disk, the spectral shape in the extreme ultraviolet and how this connects to the soft X-ray is largely unknown due to absorption by the Milky Way ISM. At intermediate redshifts, portions of this band become directly visible. Temperatures producing a thermal peak at ∼1200 Å (Telfer et al. 2002) are too cool for thermal radiation from the accretion disk to continue to the soft X-ray band (e.g., Done et al. 2012), and a likely explanation for the extreme ultraviolet continuum is Comptonization of the disk spectrum by a warm, ionized coronal layer just above the disk or near its inner edge, as shown in Fig. 3.

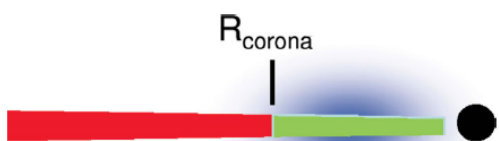

**Figure 3:** The thermally emitting outer disk is in red. In the inner disk (green) thermal photons are Compton scattered by "warm" gas to make the extreme ultraviolet/soft-X-ray excess. A hot inner corona (blue) Compton scatters the thermal radiation from the outer disk to produce the hard X-ray power law. (Adapted from Done et al. 2012.)

Jin et al. 2012 fit such a 3-component model to the SEDs of a sample of 51 low-z AGN with SDSS and XMM-Newton spectra. Fig. 4 shows striking variations in the contribution of the disk thermal component and the warm Comptonized contribution in the unobservable extreme UV. Observation of this portion of the spectrum in intermediate redshift AGN (z~1) will directly reveal objects with dominant thermal peaks (e.g., RBS 0769 in Fig. 4) and those with strong Comptonized tails (like PG1114+407). Existing ground-based observations (e.g., SDSS DR7, Shen et al. 2011) would give fundamental parameters such as $M_{BH}$ and $L_{Edd}$ to test the hypothesis that strong thermal peaks at short wavelengths are correlated with super-Eddington accretion, while strong Comptonized tails predominate in sub-critical disks. In objects with a strong Comptonization component (like PG1114+407 in Fig. 4), simultaneous optical observations would allow direct correlation of the soft seed photons from the disk with the Compton-scattered EUV. The high sensitivies we discuss below would allow high S/N observations on the short intraday timescales expected for radiative reprocessing in the disk. Lags in the correlation would then yield the geometry of the scattering region.

**Figure 4:** Three-component continuum fits to three example AGN from Jin et al. (2012). Left: RBS 0769 ($L/L_{Edd}$=13). Center: PG1004+130 ($L/L_{Edd}$=0.08). Right: PG1115+407 ($L/L_{Edd}$=0.4).

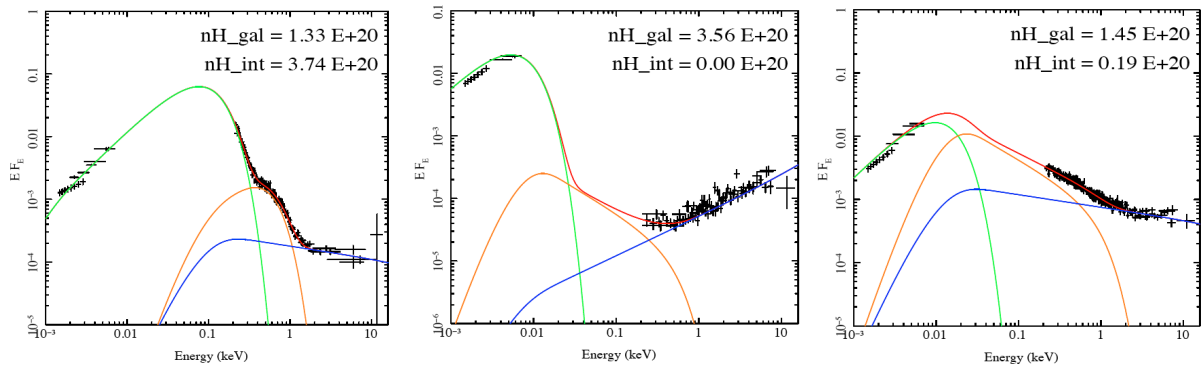

As in the prior two examples discussed above, the same sample of AGN used as background sources for probing the ISM, the IGM and the CGM would simultaneously provide the necessary targets for the scientific objectives described here.

### **Direct Black Hole Mass Measurements to Cosmological Distances**

Understanding the influence of black holes on galaxy formation and evolution also requires knowledge of galaxy structures and black hole masses as a function of redshift. Direct dynamical measurement of black hole masses requires only modest spectral resolution but high spatial resolution. Angular scales of 0.01" resolve nuclear gas disks similar to that in M87 (Ford et al. 1994) under the influence of central black holes with $M_{BH} > 10^{9.3}$ $M_\odot$ at all redshifts (Batcheldor & Koekemoer 2009). Keplerian velocities of hundreds of km s$^{-1}$ require resolving powers of only ~1000. Surface brightness dimming then becomes the limiting factor, requiring a corresponding increase in telescope aperture. Batcheldor & Koekemoer (2009) show that the resolution and low sky brightness afforded by the Lyα emission line in the UV is more efficient than 30-m ground-based telescopes in the IR. A disk with the Lyα surface brightness of M87 requires an 8-m space-based telescope to make the required observations to a limiting redshift of z=1.5 (see Fig. 5). Equipping such a telescope with an integral-field spectrograph with a field of view of ~1" and 4 mas pixels would achieve such measurements to a S/N of 5-10 in exposure times of 17—70 h.

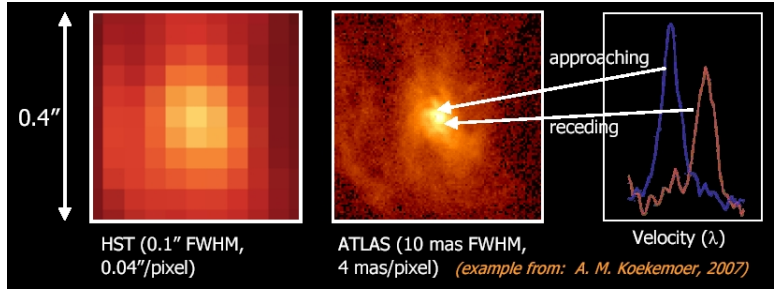

**Figure 5:** An example of a 250 pc radius nuclear disk (based on the M87 black hole, Ford et al. 1994), extrapolated to z~1.5 (where 250 pc subtends ~0.03"). (Left) Appearance of the disk with HST; (Middle) Appearance of the disk with a ~10 mas PSF (8-m space telescope, diffraction limited to 3000 Å); (Right) spectra in the central few hundred parsecs.

## Observational Requirements for AGN Science

Spectral features in the absorption troughs of AGN outflows typically show component widths of 30—100 km s$^{-1}$. Resolving these in the lithium-like doublets to decompose the column density and covering fraction requires resolving powers of R~15,000. Measuring the column density and covering fraction in deep absorption troughs requires S/N=10 *in the bottom of the trough*. For typical troughs that are 10% of the continuum level, this implies S/N~30 in the continuum. COS can reach this S/N at flux levels of $F_\lambda > 6\times10^{-14}$ erg cm$^{-2}$ s$^{-1}$ Å$^{-1}$ in 1 orbit, or ~2000 s. This is equivalent to i=13.5 for a QSO with an SED matching the SDSS composite spectrum. However, only a handful of AGN are this bright, i.e., all the examples cited in the preceding sections. The scientific objectives described above require comprehensive surveys providing comparable spectra of hundreds of AGN. Short observing times not only enable large surveys, but also permit us to probe intraday variability. At i<17, and predicted $F_\lambda > 1\times10^{-15}$ erg cm$^{-2}$ s$^{-1}$ Å$^{-1}$, SDSS DR7 has over 250 AGN with 0.89 < z < 1.50 (which enables us to see Mg X at λ>1150 Å and Lyα at λ<3200 Å). Observing these targets in 2000 s requires ~60x the throughput of COS. All requirements can be readily met with an 8-m UV/optical telescope with sensitivity down to the Lyman limit at 912 Å.

The lead submitter, Gerard Kriss, would be happy to present the science objectives described here at a NASA workshop.

**References:**
Arav, N., et al.  1999ApJ...516...27A
Arav, N., et al.  2007ApJ...658..829A
Arav, N., et al.  2012, astro-ph
Baldry, I., et al.  2004ApJ...600..681B
Batcheldor, D., & Koekemoer, A. M. 2009PASP..121.1245B
Cole, S., et al.  2001MNRAS.326..255C
Crenshaw, D. M. et al.  2003ARA&A..41..117C
Crenshaw, D. M., & Kraemer, S. B. 2012ApJ...753...75C
DiMatteo, T., et al. 2005Natur.433..604D
Done, C., et al. 2012MNRAS.420.1848D
Elvis, M.  1994ApJS...95....1E
Ford, H. C. et al. 1994ApJ...435L..27F
Gabel, J. R., et al.  2005ApJ...631..741G
Hamann, F.  1997ApJ...478...80H
Hopkins, A. M., & Beacom, J. F. 2006ApJ...651..142H
Hopkins, P. F., & Elvis, M.  2010MNRAS.401....7H
Huang, J.-S., et al. 2003ApJ...584..203H
Jin, C., et al. 2012MNRAS.420.1825J
Kraemer, S. B., et al. 2006, 2006ApJS..167..161K
Muzahid, S. 2012MNRAS.424L..59M
Shang et al., Z.  2011ApJS..196....2S
Shen, Y., et al. 2011ApJS..194...45S
Telfer, R., et al.  1998ApJ...509..132T
Telfer, R., et al.   2002ApJ...565..773T
Zheng, W., et al.  1997ApJ...475..469Z